\documentclass[aps,prb,twocolumn,showpacs,preprintnumbers]{revtex4}
\usepackage{graphicx}
\begin{document}

\title{Influence of strong microwave radiation on static properties
of a small Josephson junction}
\author{I. Abal'osheva, P. Gier\l owski, M. Jaworski, and S. J. Lewandowski}
\affiliation{Institute of Physics, Polish Academy of Sciences, Al.
Lotnik\'ow 32/46, 02-668 Warszawa, Poland}

\begin{abstract}
Preliminary results are presented concerning static properties of a
small Josephson junction under the influence of strong microwave
radiation. We discuss the correspondence between a Brownian particle
moving in a periodic potential and superconducting phase difference
in a small Josephson junction. Next, we describe an experimental
method of determining the amplitude of  microwave current flowing
across the junction. Typical examples of static characteristics of
the junction are presented, including its dynamical resistance as a
function of microwave power. We discuss also the influence of an
external magnetic field on the junction dynamics and show that in
this case the one-dimensional Stewart-McCumber model becomes
insufficient.
\end{abstract}

 \maketitle

\section{Introduction}

Recently, the noise-assisted transport of Brownian particles has
attracted considerable attention, which is still growing in view of
possible applications in various branches of physics and
chemistry.\cite{HB,AstH,RH} Unfortunately, experimental
investigation of transport phenomena on a molecular scale is
extremely difficult, thus apart from theoretical studies there is a
need for adequate physical systems, modeling various transport
phenomena in a way which could be accessible experimentally.

As shown in Refs.~\onlinecite{LPRL,LPRB}, there is a rigorous
mathematical equivalence between the Newton-Langevin (NL) equation,
describing a Brownian particle in a periodic potential, and the
Stewart-McCumber (SM) model dealing with the superconducting phase
dynamics in a small Josephson junction in the presence of thermal
fluctuations. In particular, a constant external force acting on a
Brownian particle corresponds to a constant current biasing the
Josephson junction, while a time-averaged velocity of the particle
is equivalent to a constant voltage induced on the junction
electrodes. In other words, the relevant parameters of a Brownian
motor can be related to easily measurable macroscopic quantities
such as a constant current and constant voltage, leading directly to
the current-voltage ($I$-$V$) characteristic of the Josephson
junction.

The aim of the present paper is twofold. First,  we propose a
uniform method making possible to determine the absolute value of
microwave current amplitude between junction electrodes,
independently of the junction geometry and its coupling to the
experimental setup. Such a uniform approach covers a wide range of
junction parameters, including two limiting cases of a ``current
source'' and a ``voltage source'' model. Second, we investigate the
influence of a strong microwave signal on the static properties of a
small Josephson junction, with particular attention paid to its
dynamic (differential) resistance. In this context we present
preliminary results and explore the possibility of detecting a
negative resistance (both absolute and differential) on the $I$-$V$
characteristic.

The paper is organized as follows. In Sec. II we discuss the
equivalence between the NL and SM models. Next, we recall the
well-known relation between the critical current $I_0$ and the
normal resistance $R_n$ of a real Josephson junction, and discuss
the possibility of making the junction parameters compatible with
theoretical estimates. Experimental details, including  coupling of
the junction to a microwave setup, are given in Sec. III. Sec. IV
deals with an experimental evaluation of  rf (microwave) current
flowing across the junction. In particular, we propose an analytical
approximation making possible to determine the rf current amplitude
for a wide range of junction parameters. A few examples of the
current-voltage ($I$-$V$) characteristics are presented in Sec. V
with a special attention paid to the dynamic resistance as a
function of strong microwave signal. We discuss also the influence
of an external magnetic field on the junction behavior and show that
the superconducting phase difference $\phi$ becomes spatially
modulated, making the SM model insufficient for a proper description
of the junction dynamics. Sec. VI contains concluding remarks, we
indicate also possible generalization to a multidimensional model of
a Josephson junction immersed in an external magnetic field.

\section{Modeling of chaotic phenomena in a Josephson junction}

The chaotic dynamics of a classical Brownian particle in the
presence of thermal noise can be described by the inertial NL
equation:\cite{LPRL}
\begin{equation}
m\ddot{x}+\gamma\dot{x}+V^\prime(x)=F+A\cos{\Omega t}+ \sqrt{2\gamma
k_BT}\xi(t)
\end{equation}
where a dot and prime denote differentiation with respect to $t$ and
$x$, respectively. The particle mass is denoted by $m$, $\gamma$ is
the friction coefficient, $\Omega$ --- the angular frequency, $F$
and $A$ denote constant and alternate external forces, respectively.
Thermal fluctuations are described by zero-mean Gaussian white noise
$\xi(t)$ with the autocorrelation function
$\langle\xi(t)\xi(u)\rangle=\delta(t-u)$, $k_B$ denotes the
Boltzmann constant and $T$ --- temperature.

If a spatially periodic potential  $V(x)$ is cosinusoidal, then it
can be easily verified that Eq. (1) is formally equivalent to the SM
model describing the superconducting phase dynamics of a
one-dimensional small-area Josephson junction \cite{LPRB,Kautz}
\[
\left(\frac{\hbar}{2e}\right)C\ddot{\phi}+\left(\frac{\hbar}
{2e}\right)\frac{1}{R_n}\dot{\phi}+I_0\sin(\phi)
\]
\vspace{-5mm}
\begin{equation}
= I_d+ I_a\cos(\Omega t) +\sqrt{\frac{2k_BT}{R_n}}\xi(t)
\end{equation}
where $\phi$ denotes the superconducting phase difference between
electrodes and the junction is characterized by its critical current
$I_0$, capacitance $C$ and normal-state resistance $R_n$. The
amplitudes of the dc and ac (microwave) currents applied to the
junction are denoted by $I_d$ and $I_a$, respectively.

A dimensionless form of Eq. (2) can be written as
\begin{equation}
\frac{d^2\phi}{dt^{\prime 2}}+\sigma\frac{d\phi}{dt^\prime}
+\sin(\phi)=i_0+i_1\cos(\Omega_1t^\prime)+\sqrt{2\sigma
D}\Gamma(t^\prime)
\end{equation}
where the time has been rescaled to $t^\prime=t/\tau_0$,
$\tau_0=1/\omega_p=\sqrt{\hbar C/2eI_0}$. Consequently,
$\Omega_1=\Omega\tau_0$, $\sigma=\tau_0/R_nC$, and the noise intensity
$D=2ek_BT/\hbar I_0$. The dimensionless dc and ac amplitudes are
given by $i_0=I_d/I_0$ and $i_1=I_a/I_0$, respectively.

Comparison of Eq. (3) with Eq. (1) shows clearly that (after
appropriate scaling of Eq. (1)) both models are mathematically
equivalent, although the physical meaning of relevant variables is
distinctly different. In particular the superconducting phase
difference $\phi$ in Eq. (2) corresponds to the spatial coordinate
$x$ in Eq. (1). Consequently, the average value of the
time-derivative $\langle\phi_t\rangle$, proportional to the constant
voltage across the junction, is analogous to the averaged velocity
of a Brownian particle. Similarly, external current amplitudes $I_d$
and $I_a$ correspond to external forces $F$ and $A$, respectively.

The range of dimensionless parameters in which we can expect either
an absolute or a differential negative resistance, is given in
Ref.~~\onlinecite{LPRB}. Thus, using the scaling of Eq. (3) and
coming back to Eq. (2) it is possible to recover absolute values of
the junction parameters and the external driving.

For example, the noise intensity $D=10^{-3}$ implies $I_0\simeq 176\
\mu$A at $T=4.2$ K. Moreover, assuming $\Omega=150\times 10^9$, i.e.
$f\simeq 23.87$ GHz we obtain the following sets of junction
parameters
\begin{eqnarray}
{\rm (a)}\ \ R_n=2.6\ \Omega, &C=13.8\ {\rm pF}, &I_0=176\ \mu{\rm A}\nonumber\\
&&\\
{\rm (b)}\ \ R_n=1.4\ \Omega, &C=27.9\ {\rm pF}, &I_0=176\ \mu{\rm
A} \nonumber
\end{eqnarray}

The external rf current amplitude may vary, ranging from $I_a\simeq
120\ \mu$A for the set (a) up to $I_a\simeq 500\ \mu$A for the set
(b).\cite{LPRB}

The main difficulty with practical application of the above
parameters follows from the fact that for a real Josephson junction
the normal resistance $R_n$ and the critical current $I_0$ are not
independent, but their product is given by:\cite{BP}
\begin{equation}
I_0R_n=\frac{\pi\Delta(T)}{2e}\tanh\left(\frac{\Delta(T)}{2k_BT}\right)
\end{equation}
where $\Delta(T)$ denotes the superconducting gap energy.

For niobium at $T=4.2$ K, we have $\Delta(T)\simeq 1.4\ $meV, thus
the upper limit for $I_0R_n$ can be estimated according to Eq. (5)
as $I_0R_n\simeq 2\ $mV. In practice, the value of $I_0R_n$ for a
real junction is lower due to inevitable defects, nevertheless it is
still a few times greater than the value of $I_0R_n$ resulting from
the set of parameters (4) required by the theory.

In order to make real junction parameters compatible with
theoretical predictions one can use one of two experimentally
accessible methods, reducing either $R_n$ or $I_0$. In the first
case, the effective value of normal resistance can be decreased by
shunting the junction with an additional resistor connected in
parallel to the junction electrodes.\cite{Nagel} In this work,
however, we use an alterative approach to reduce the critical
current $I_0$ by applying an external constant magnetic field in the
junction plane.

The following sets of junction parameters have been chosen for
further studies:
\begin{eqnarray}
{\rm (a)}\ \ R_n=2.96\ \Omega, &C=14.0\ {\rm pF}, &I_0=460\ \mu{\rm A}\nonumber\\
&&\\
{\rm (b)}\ \ R_n=1.45\ \Omega, &C=28.0\ {\rm pF}, &I_0=920\ \mu{\rm
A} \nonumber
\end{eqnarray}

Comparing the above parameters with theoretical predictions (4) one
can see that the values of $R_n$ and $C$ correspond approximately to
those specified previously, while the nominal critical currents
$I_0$ are a few times greater than $I_0\simeq 176\ \mu$A required by
the theory. As mentioned above, the critical current can be adjusted
by applying an external magnetic field to the junction. In this
manner, we are able to obtain the relevant junction parameters in
agreement with those predicted theoretically.

The amplitude $I_a$ of the rf current flowing across the junction is
obviously proportional to $\sqrt{P}$, where $P$ denotes the incident
microwave power. Unfortunately, $I_a$ depends in a rather
complicated way on the geometry of the sample holder and its
coupling to the microwave source. Therefore, the absolute value of
$I_a$ has been determined experimentally by using the well-known
relations between the effective constant current at zero voltage and
the rf current amplitude $I_a$ (see Section IV for details).

\section{Experimental}

The Nb-AlO$_x$-Nb Josephson junctions in overlap geometry have been
designed and manufactured  in the Institute of Radio Engineering and
Electronics in Moscow. The sample holder was designed as a section
of a ridge waveguide of reduced impedance, which was used to improve
matching of the junction to the standard $K$-band (WR 42) waveguide
and obtain a relatively large amplitude $I_a$ of the rf current.
Additionally, $I_a$ could be controlled by a variable short located
just behind the sample holder. Two superconducting coils of nearly
rectangular cross-section were attached to the side walls of the
sample holder, and could be used to apply a dc magnetic field of up
to $\sim 40$ Gs directed parallel to the junction plane.

The measurement system consisted of a $K$-band klystron, followed by
a ferrite isolator, ferrite modulator and calibrated attenuator. The
microwave frequency could be controlled in the range 23 -- 25 GHz
and the maximum output power  level was estimated as $100$ mW.

The current-voltage ($I$-$V$) characteristics were measured by a
standard 4-point technique. The junction was biased by a digital
current source while the voltage was measured using a digital
voltmeter and recorded by a computer. All the measurements were
carried out at the temperature $T=4.2$ K.

\section{Evaluation of the microwave current amplitude}

According to the ``current source'' model \cite{Rich} we consider
the following equation:
\begin{equation}
\frac{d\phi}{d\tau}+\sin(\phi)=i_0+i_1\sin(\xi\tau),
\end{equation}
where $i_0=I_d/I_0$, $i_1=I_a/I_0$ as before, $\xi=\Omega/\omega_c$
and $\omega_c=2eR_nI_0/\hbar$. As compared to Eq. (3), we use a
slightly different time scaling $\tau=\Omega_c t$ and neglect both
the second time-derivative and the noise term.

Eq. (7) has been solved numerically in several papers
\cite{BP,Rich,Russ,JJAP} and it has been shown that the maximum
constant current $I_{max}$ at zero voltage decays with increasing rf
amplitude $i_1$. However, detailed behavior of $I_{max}$ as a
function of $i_1$ depends strongly on the frequency parameter $\xi$.
For $\xi\ll 1$, we have a ``current source'' model, $I_{max}$
decreases linearly and $I_{max}=0$ is attained for $i_1\simeq 1$. On
the other hand, for $\xi\gg 1$ the model tends to the ``voltage
source'' with a Bessel-like behavior:
\begin{equation}
I_{max}/I_0=J_0\left(\frac{2eR_nI_a}{\hbar\Omega}\right),
\end{equation}
where $J_0$ denotes the Bessel function of order 0 and the argument
can be transformed to a simpler form by using dimensionless
quantities:
\begin{equation}
\frac{2eR_nI_a}{\hbar\Omega}=\frac{\omega_ci_1}{\Omega}=\frac{i_1}{\xi}.
\end{equation}

\begin{figure}
\includegraphics[scale=0.35]{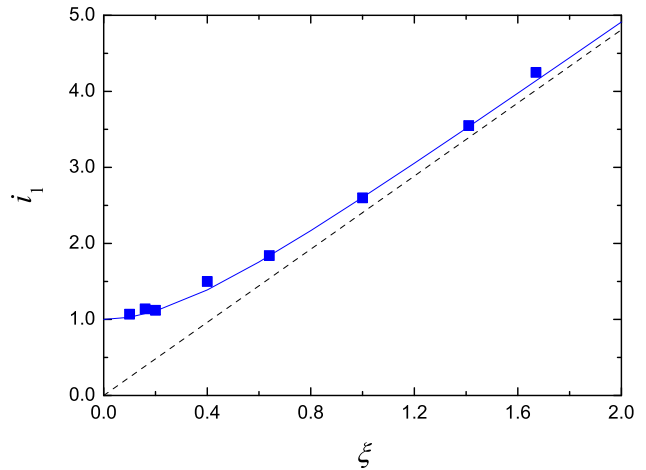}
\caption{(Color online) The relative rf current amplitude $i_1$ at
$I_{max}=0$ plotted as a function of the normalized frequency $\xi$.
Full squares denote numerical results, the dashed line shows an
asymptotic linear dependence, and the solid line represents an
analytical approximation (10).}
\end{figure}

Hence, asymptotically $I_{max}=0$ is obtained at
$i_1=\kappa_{0,1}\xi$, where $\kappa_{0,1}=2.405$ denotes the first
zero of the Bessel function $J_0$. Fig. 1 shows the results of
numerical calculations of $i_1|_{I_{max}=0}$ as a function of $\xi$,
taken from Refs.~\onlinecite{BP,Rich,Russ,JJAP}. The dashed line
denotes an asymptotic linear dependence $i_1=\kappa_{0,1}\xi$ and
the solid line shows an analytical approximation suggested here to
cover the whole range of $\xi$:
\begin{equation}
i_1=\sqrt{1+(\kappa_{0,1}\xi)^2}.
\end{equation}

It is clear that the above approximation exhibits a proper behavior
both for $\xi\rightarrow 0$ and $\xi\gg 1$. Moreover, as shown in
Fig. 1, the analytical approximation (10) is in good agreement with
numerical data, thus it can be used to evaluate the rf current
amplitude $i_1$ also for intermediate values of $\xi$.

\begin{figure}
\includegraphics[scale=0.35]{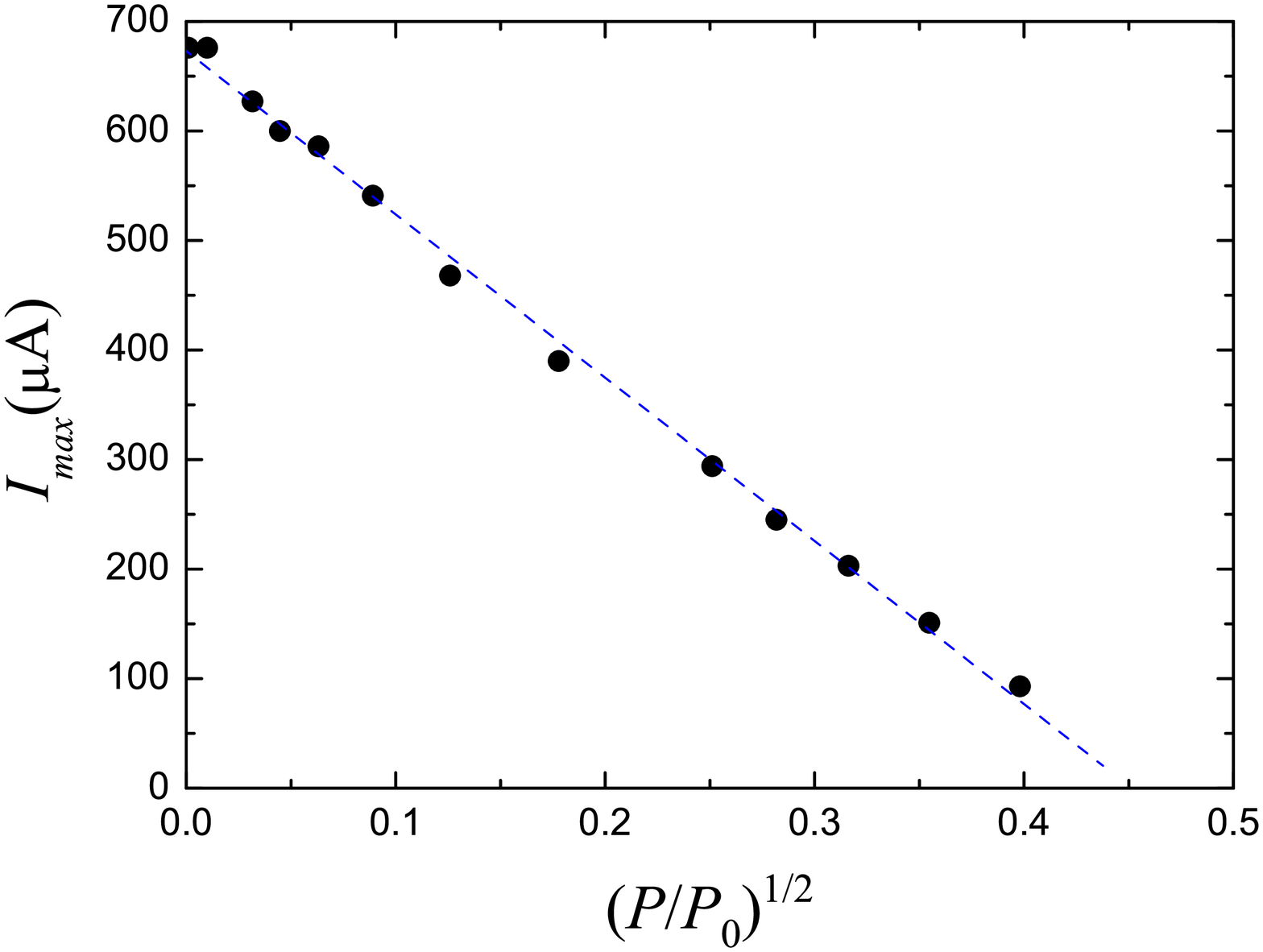}
\caption{(Color online) Experimental dependence of the maximum
constant current $I_{max}$ (full circles) on the relative amplitude
of the microwave signal $(P/P_0)^{1/2}$. The dashed line shows a
linear fit to the experimental data.}
\end{figure}

As an example, Fig. 2 shows an experimental dependence of $I_{max}$
on the relative amplitude of microwave signal $(P/P_0)^{1/2}$, where
$P/P_0$ denotes the relative microwave power.

The junction parameters are slightly different from those given in
Eq. (6) and are equal to $R_n=2\ \Omega$, $I_0=680\ \mu$A, hence
$\xi\simeq 0.037$ for the microwave frequency $f\simeq 24$ GHz, and
according to the analytical approximation (10) we find
$i_1|_{I_{max}=0}\simeq 1.004$. It follows from Fig. 2 that
$I_{max}=0$ for $(P/P_0)^{1/2}\simeq 0.45$ what means that the
absolute amplitude of the rf current at this point is equal to
$I_a\simeq I_0=680\ \mu$A. Thus, the maximum available current
amplitude can be evaluated as $I_a^{(max)}=I_a/0.45\simeq 1.5$ mA and is
far above the range of $I_a$ predicted by theoretical considerations
(see Section II). Strictly speaking, the above analysis deals with
the RSJ model, since the second time-derivative has been neglected
in Eq. (7). The influence of the junction capacitance on the rf
current amplitude is not fully understood, however as follows from
Ref.~\onlinecite{JJAP}, the estimates taking into account a nonzero
capacitance yield even larger values of $I_a$.

To conclude this section, we were able to estimate the rf current
amplitude for a wide range of dimensionless parameter $\xi$, which
depends on the junction parameters and the rf frequency. In
particular, it was shown that the rf amplitudes required by the
theory could be easily obtained in our experimental setup.

\section{$I$-$V$ measurements --- results and discussion}

In order to study the influence of microwave radiation on the static
properties of a Josephson junction, we have performed a series of
detailed $I$-$V$ measurements for various levels of microwave power.
Twelve sets of junctions have been measured with the parameters
close to those given in Section II.

\begin{figure}
\includegraphics[scale=0.35]{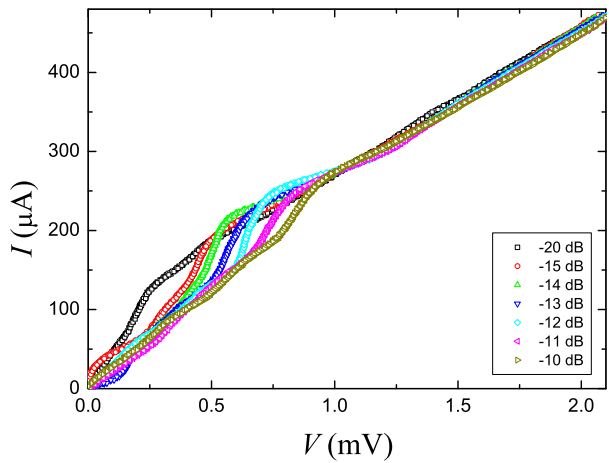}
\caption{(Color online) Family of $I$-$V$ characteristics plotted as
a function of incident microwave power level.}
\end{figure}

The results obtained so far exhibit a rich variety of different
phenomena which are still under investigation. Here we present only
a few examples, dealing mainly with the dynamical resistance of the
junction in the presence of strong microwave radiation.

\begin{figure}
\includegraphics[scale=0.35]{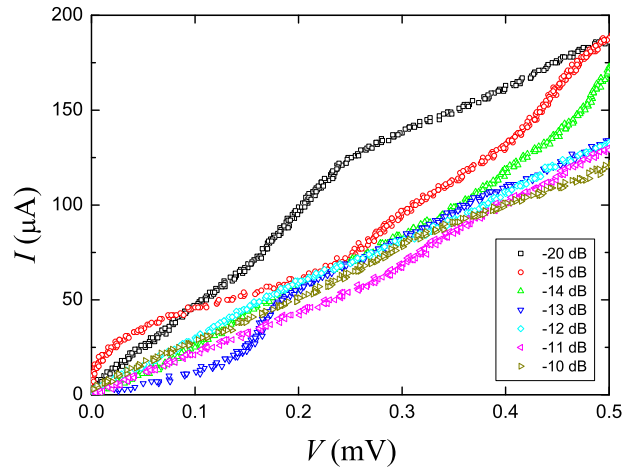}
\caption{(Color online) Enlarged part of Fig. 3 plotted for smaller
values of current and voltage.}
\end{figure}

A typical example of the $I$-$V$ characteristic measured for the
sample (a) is presented in Fig. 3, while Fig. 4 shows an enlarged
part of the same diagram plotted for smaller current and voltage
values. According to Section IV, the power level $P/P_0=-20$ dB
corresponds to the absolute rf amplitude $I_a\simeq 150\ \mu$A and
consequently $-10$ dB denotes $I_a\simeq 480\ \mu$A. For the set of
parameters given by Eq. (4), we have $\Omega_1<1$ and
$\Omega_1>\sigma$. As a consequence, the Shapiro steps are not
observed in this region, in agreement with
Refs.~\onlinecite{LPRB,Kautz}.

It is clear that the $I$-$V$ characteristics are influenced strongly
by the microwave radiation. First, we observe a step in the $I$-$V$
curve, which corresponds to a local decrease of the dynamical
resistance $R_d=dV/dI$ from an asymptotic value $R_d\simeq 5.3\
\Omega$ to $R_d\simeq 1.1\ \Omega - 1.5\ \Omega$. The voltage
position of this step, denoted by $V_s$, depends linearly on the rf
amplitude as shown in Fig. 5.

\begin{figure}
\includegraphics[scale=0.35]{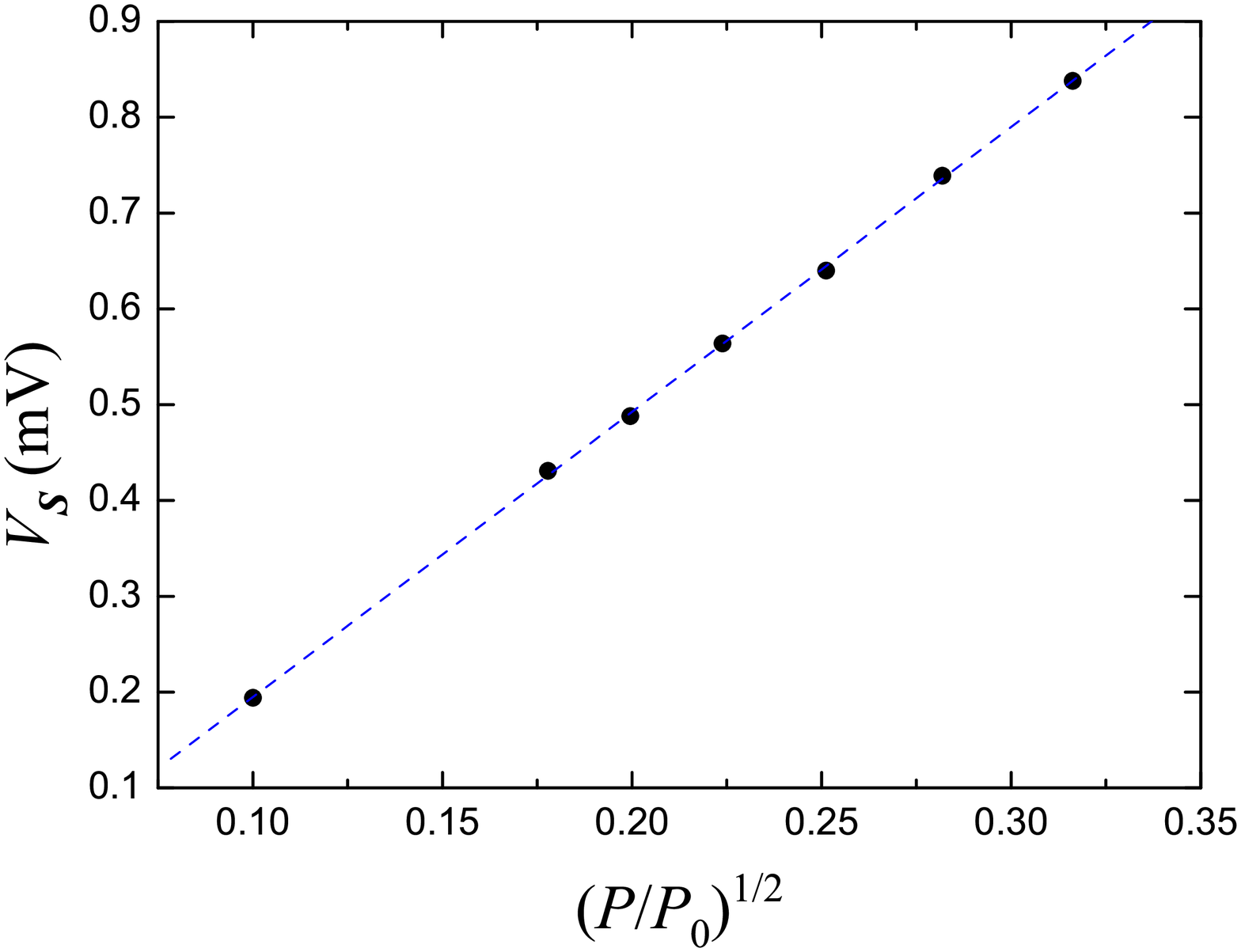}
\caption{(Color online) Step position $V_s$ as a function of the
relative rf amplitude. Experimental data are plotted as full circles
and the dashed line shows a linear fit.}
\end{figure}

The second effect of the microwave
radiation is its influence on the dynamical resistance $R_d$ at zero
voltage (see Fig. 6). This time we observe rather large and
irregular changes of $R_d$ ranging from a very small value $R_d<0.1\
\Omega$ for $(P/P_0)^{1/2}=0.177$ up to $R_d\simeq 8.4\ \Omega$ for
$(P/P_0)^{1/2}=0.224$. It is clear that the dependence of $R_d$ on
the rf radiation is non-monotonic and for $(P/P_0)^{1/2}=0.177$ we
observe nearly zero value of the dynamical resistance.
Unfortunately, contrary to theoretical predictions, so far we were
not able to observe a negative dynamical resistance of the Josephson
junction.

\begin{figure}
\includegraphics[scale=0.35]{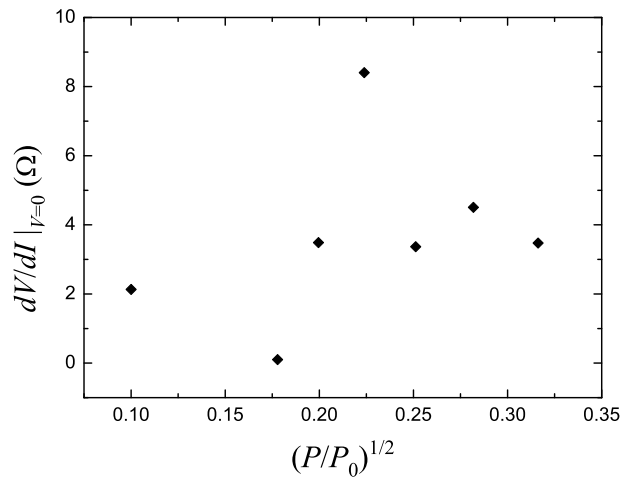}
\caption{(Color online) Dynamical resistance at zero voltage as a
function of the relative rf amplitude.}
\end{figure}

It seems that the main reason for a visible discrepancy between
theoretical and experimental results follows from the method used to
reduce the critical current of the junction. A more detailed
analysis shows \cite{VD} that applying an external magnetic field in
the junction plane gives rise to a spatial modulation of the current
density. In other words, the external magnetic field makes the phase
$\phi$ to become a function of spatial coordinates, in spite of the
fact that we deal with small junctions whose physical dimensions are
much smaller than the Josephson penetration depth $\lambda_J$. As a
consequence, the Stewart-McCumber model becomes inadequate and
should be generalized to a multidimensional case, taking the form of
a partial differential (sine-Gordon like) equation:
\begin{equation}
-\alpha\nabla^2\phi+\frac{d^2\phi}{dt^2}+\sigma\frac{d\phi}{dt}
+\sin(\phi)=i_0+i_1\cos(\Omega t)+\sqrt{2\sigma D}\Gamma(t)
\end{equation}
where $\nabla^2\phi=d^2\phi/dx^2+d^2\phi/dy^2$ denotes the
two-dimensional Laplacian, $\alpha$ is a scaling parameter and
$\phi(x,y,t)$ is a function of time and two spatial coordinates.

Unfortunately, such a model has no counterpart in the domain of
chaotic dynamical systems, hence the theoretical predictions
resulting from the Newton-Langevin equation cannot be directly
applied to a Josephson junction immersed in an external magnetic
field.

\section{Summary and conclusions}

In this work we investigated the influence of high-power microwave
radiation on the static characteristics of a small Josephson
junction. In particular, the dependence of the critical current on
the microwave power level was studied, allowing to determine the
absolute value of the rf (microwave) current flowing across the
junction. It was shown that for the junction parameters as given in
Section II, a ``current source'' model can be applied, leading to a
simple relation $I_a\simeq I_0$ at the point where the maximum
constant current $I_{max}$ attains zero.

In the next step, a dynamical resistance of the junction was
investigated as a function of the rf current amplitude. In
particular, it was found that the dynamical resistance at zero
voltage is strongly dependent on the microwave power level and may
vary in a wide range, however negative values of the resistance were
not observed.

As far as the correspondence between Newton-Langevin and
Stewart-McCumber models is concerned, we found that the junction
parameters required by the theory were incompatible with those of
really existing Josephson junctions due to a natural constraint on
the product of $R_n$ and $I_0$. As pointed out in Section V, an
external magnetic field applied in the junction plane may reduce the
critical current, but also makes the superconducting current density
to be spatially modulated. In other words, the analogy between the
NL and SM models breaks down, since the dynamics of a Josephson
junction immersed in an external magnetic field should be described
by a more general multidimensional equation.

\section*{Acknowledgments}

This work was supported by the Grant No. N202 203534. Thanks are
also due to Prof. V. P. Koshelets for making the Josephson junctions
available.

\end{document}